# Observation of optical filtering effects with four-wave mixing in a cold atomic ensemble

Dong-Sheng Ding[1,2,#], Yun Kun Jiang[3,#], Wei Zhang[1,2], Zhi-Yuan Zhou[1,2], Bao-Sen Shi[1,2,*], and Guang-Can Guo[1,2]

[1]*Key Laboratory of Quantum Information, University of Science and Technology of China, Hefei 230026, China*
[2]*Synergetic Innovation Center of Quantum Information & Quantum Physics, University of Science and Technology of China, Hefei, Anhui 230026, China*
[3]*College of Physics and Information Engineering, Fuzhou University, Fuzhou, 350002, P. R. China*
[*]*Corresponding author:* [*]drshi@ustc.edu.cn
[#]These authors contribute this article equally.

We observe an optical filtering effect in four-wave mixing (FWM) process based on a cold atomic gas. The side peaks appear at the edges of pulse of generated optical field, and they propagate through the atomic media without absorption. The theoretical analysis gives that these side peaks corresponded to the high frequency part of pulse of generated signal, which means the atoms cannot response to the rapid change of the electromagnetic field in time. On the contrary, the low frequency components of generated signal are absorbed during the transmission through the atoms. In addition, we experimentally demonstrate that the backward side peak could be stored by using Raman transition in atomic ensemble and retrieved later.

There is an interesting optical phenomenon that the edge of a classical electromagnetic field pulse directly goes out when it propagates through a absorbing media with an index of refraction $n(\omega)$, which is called an optical precursor [1, 2]. This phenomenon is studied in gamma ray [3], microwave [4], optical regimes [5-8] and in electromagnetically induced transparency (EIT) media [9-15]. The underlying physical mechanics is that the response of the media is too late to follow the high frequency components of electromagnetic field. In EIT media, Du et al [9, 12-15] have experimentally observed the precursor with a classical electromagnetic field [12] or a single photon [14] respectively, in which the low frequency components of input signal field are delayed by atoms and the high frequency components go through the atoms without any delay.

There are no any experimental demonstrations of optical response of atoms in a nonlinear process, such as Four-wave mixing (FWM) process. FWM has been studied by many groups in an atomic system with the configuration of a double lambda [16-19], a ladder or a diamond type [20-24]. Usually, FWM process is used to realize the frequency conversion of a photon or an optical pulse. In this paper, we reported on the observation of an interesting optical filtering effect in a FWM process with an inverted-Y type configuration in cold atoms: the optical side peaks of generated signal field appeared. We theoretically found and experimentally proved that: if the edge of input pulse was much steep, the side peaks were more easily generated. In addition, we experimentally realized the storage of the backward side peak, which showed that the off-resonance storage could be realized even if the optical density is low. Our results gave a new understanding in nonlinear optics and optical storage: the responding rate of atoms in the nonlinear process was always larger than the linear process in an ensemble media.

The inverted-Y type configuration used in our experiment was shown in Fig. 1(a). It consisted of two degenerated ground states |1> and |2> ($5S_{1/2}$ F=3), one intermediate state |3> ($5P_{1/2}$ F'=2, decay rate $\gamma$) and one upper state |4> ($4D_{3/2}$ F''=2, decay rate $\Gamma$). The atomic transition between the ground state and the intermediate state matched to D2 line of Rubidium 85 ($^{85}$Rb), and the transition between the intermediate state and the upper state could be driven by another laser at a wavelength of 1475.6 nm. We experimentally generated a new electromagnetic field at 795 nm by combining coupling laser, pump 1 and probe fields via a non-collinear FWM configuration.

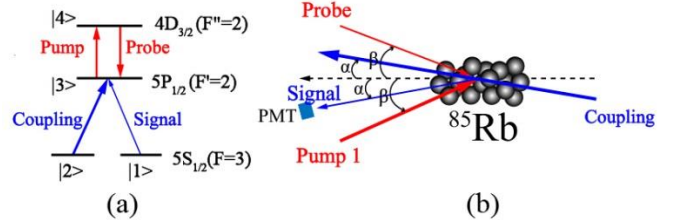

Fig. 1 The experimental energy level diagram was shown by (a). The two ground degenerated states were the sublevels of ($5S_{1/2}$ F=3) $m_F$=-1 ($m_F$=0) and $m_F$=-3($m_F$=-2). (b) Simplified experimental setup. The angles $\alpha$≈1°, $\beta$≈2°.

The simplified experimental setup was shown in Fig.

1(b). A continuous wave (cw) laser beam at 795 nm from an external-cavity diode laser (DL100, Toptica) was input into a two-dimensional magneto-optical trap (MOT) as the coupling light. A cw laser beam at 1475.6 nm from another external-cavity diode laser (DL100, Prodesign, Toptica) was divided into two beams by a beam splitter to prepare the pump and probe beams. These fields were horizontal linearly polarization and were modulated by two acousto-optic modulators. The coupling field acted as a pumping light in FWM process and as a control light in the storage process. The probe field was focused by a lens with a focus length of 500 *mm*, its Fourier plane was at the center of the atomic ensemble. The generated signal field focused by another lens was collected by a multimode fiber, and was monitored by a photomultiplier tube (PMT) (Hamamatsu, H10721).

Before introducing our experimental results, we gave a simplified theoretical description for our system. According to ref. 16, we derived steady-state solutions for the density matrix and obtained the susceptibility of signal field $\chi^{(3)}$ using equation $P_s=N\mu_{31}\rho_{31}=\varepsilon_0\chi^{(3)}E_{pr}^*$, where $N$ is the effective density of the atoms, $\mu_{31}$ and $\rho_{31}$ are the dipole element and the density matrix element of the atomic transition $|3>->|1>$ respectively. $\varepsilon_0$ is the permittivity of vacuum, $E_{pr}^*$ is the conjugate of probe field.

$$\chi^{(3)} = \frac{N\mu_{31}\mu_{43}}{i\varepsilon_0\hbar}\frac{A_1\Omega_P\Omega_C}{A_2(\Gamma_1+\Gamma)+A_3\Gamma_1}, \quad (1)$$

where $A_1$, $A_2$ and $A_3$ are complex coefficients which are the functions of pump and coupling Rabi frequencies $\Omega_P$ $\Omega_C$, decay rates $\gamma$, $\Gamma$ and the probe field detuning $\Delta_1$. The coefficient $\Gamma_1$ satisfies $\Gamma_1 = \gamma - i\Delta_2$, where the parameter $\Delta_2$ is the detuning between the atomic transition $|3>->|1>$ and signal field. We plotted the $\chi^{(3)}$ susceptibility against the detuning $\Delta_2$ with an assumption of $\Delta_1 = 0$. The result was shown by the red line of Fig. 2(a). The bandwidth of FWM process was ~ 10 MHz. The blue line was the transmission of signal field with the function of $E_s \propto E^{-\alpha/[1+(\Delta_2/\gamma)^2]}$, where the parameter $\alpha$ characterized optical density of atoms. We made the detuning of probe field $\Delta_1$ to be a variable, which was defined as $\Delta_1 = -\Delta_2$ giving the FWM process occurred under near zero two-photon detuning. We plotted the spectrum of the susceptibility $\chi^{(3)}$ given by Fig. 2(b) which is broader than the absorbing spectrum of probe.

Next, we used a square shaped pulse as an input probe shown by Fig. 3(a), the edge gradient of this pulse was characterized by a coefficient $k$ with a formula of $E_s \propto E^{-k(t-t_0)^2}$. The Fourier transformation of this pulse was given by Fig. 3(b). Through filtering with the atomic nonlinear susceptibility $\chi^{(3)}$ and the transmission spectrum of Fig. 2(b), the spectrum of Fig. 3(b) became to be Fig. 3(c). The obtained spectrum was the result of nonlinear process and atomic absorption process. Fig. 3(d) was the inversed Fourier transformation of Fig. 3(c), corresponding to the output signal pulse. From this result, we could get that high frequency components of input probe were converted into FWM field and transmitted out the atoms with a little absorption. This was due to the fact that the nonlinear spectrum of the susceptibility $\chi^{(3)}$ was broader than the linear absorption spectrum of atoms. The atoms were too late to respond the high frequency components of FWM field, so we could see the side peak of generated signal pulse. In another words, the generated signal field was always the components from FWM process with no any absorption response of atoms.

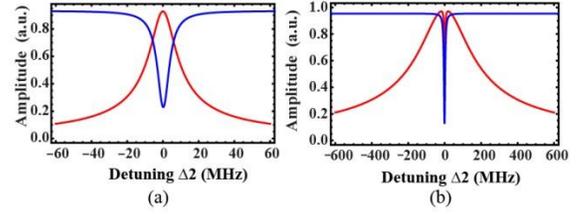

Fig. 2 (a) Red line was the spectrum of the susceptibility $\chi^{(3)}$ when the detuning of probe field was kept unchanged, blue line corresponded to the transmission spectrum of atoms when $\alpha=1$. (b) The red line is the spectrum of the susceptibility $\chi^{(3)}$ with the detuning of probe field as a variable.

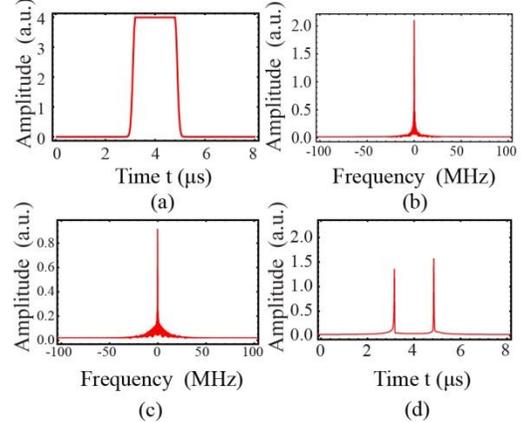

Fig. 3 (a) is the square shaped pulse of input probe, and (b) is its Fourier transformation spectrum in frequency domain. (c) is the filteringing spectrum through interactions with atoms, and (d) is the inversed Fourier transformation.

We also changed the coefficient $k$ by considering an input half-Gaussian pulse, which directly corresponded to how much the high frequency components of signal field generated. The simulated results were given by Fig. 4. In practical simulations, the timing width of the pulse $\Delta t$ was changed through arbitrary function generator AFG3252. In Fig. 4, the height of the peak decreases with the decrement of the parameter $\Delta t$. This point fully illustrated that the high frequency components of signal field was much easier to go through the atoms. In order to show this phenomenon obviously, we only made $\alpha=0.01$. For high $\alpha$, this filtering effect also appeared (see experiment in below).

We firstly performed the nonlinear FWM process using the experimental setup shown by Fig. 1(b). A cigar-shaped atomic cloud [25] was served as the nonlinear media in our experiment. We use a high efficiency camera to monitor the generated signal field. The result was shown in Fig. 5(a), where the black line was signal field, two side peaks appeared at rising and falling edge. The region between the rising and falling edge was low frequency components, strongly absorbed by atoms. As calculated before, the side peaks appeared because the spectrum of nonlinear FWM was broader than the atomic adsorption bandwidth, which resulted in the output of high frequency components of signal field.

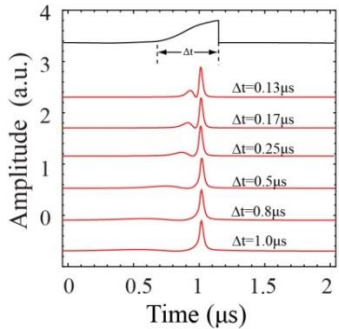

Fig. 4 The different output pulses of signal field vs the input half-Gaussian shaped pulse with different pulse width $\Delta t$.

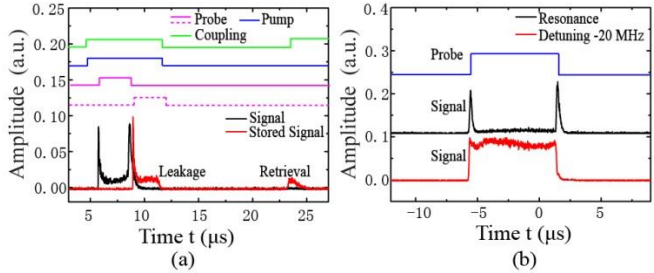

Fig. 5(a) The black line was the generated signal. The red line was the leaked and retrieved signal. The storage time was about 12 μs. The green, blue and solid purple lines were the timing sequences of coupling, pump and probe fields. The dotted purple line was the timing sequence of pump and probe fields for the purpose of storing signal. (b) The generation of signal field at resonance (black line) and with a detuning of -20 MHz (red line) respectively.

When we turned off the coupling and pump fields simultaneously, the back side peak disappeared, shown by the red line in Fig. 5(a). The FWM signal was retrieved again if we switched the coupling field on. In this process, the atomic coherence state $\rho_{12}(t)$ was prepared, resulting in memorizing signal field as spin wave of atoms. Moreover, if we used coupling laser with a detuning of -20MHz to atomic transition |3>->|1>, the side peaks didn't exist anymore, which was shown by Fig. 5(b). This was due to that the spectrum of signal field and the absorption spectrum of the atoms were not overlap.

In addition, we used a half-Gaussian pulse as a probe pulse. Fig. 6(a) was the results. The side peak appeared at the edge with large slope, there was no side peak at the edge with small slope. At the same time, we changed the width of half-Gaussian pulse to check the relation between the intensity of peak and the slope of the edge of pulse. The results were shown by Fig. 7. The side peaks appeared at the edge with large slope and disappears under small slope. These processes gave the fact that the low frequency components of pulse edge were absorbed and the high parts went through the atoms. The experimental results agreed well with the previous theoretical explanations.

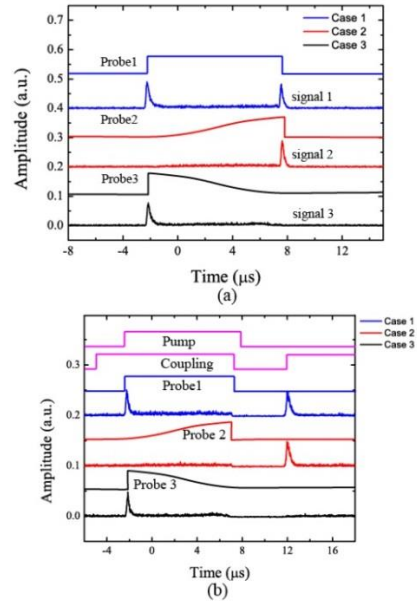

Fig. 6(a) The amplitude of generated signal field vs different shape of the probe. (b) The storage of signal field vs different shapes of input probe. Case 1 was a square-shaped pulse. Case 2 and Case 3 corresponded to a half-Gaussian pulse respectively. The powers of probe, pump and coupling are 0.45mW, 2.8mW and 0.5mW respectively.

We tried to store these generated side peaks, the results were shown by Fig. 6(b). The case 1 and 2 in Fig. 6(b) showed clearly that the generated backward side peak of signal could be stored and retrieved later. In case 3, there was no any storage of signal because there wasn't high frequency component in backward edge Certainly, our results also supported the observation in ref. 12 and 14, if the broadening frequency of pulse's edge was far more than the bandwidth of storage, there was no observation of storing narrow pulse or delaying it.

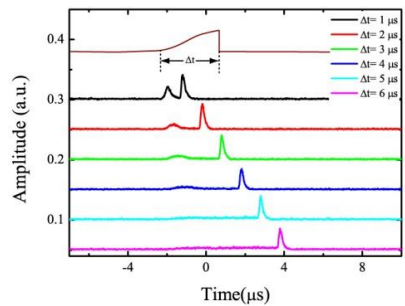

Fig. 7 The generated signal field vs different period of the half-Gaussian pulse from 1 μs~6 μs. The powers of probe, pump and coupling were 0.45mW, 2.8mW and 0.5mW.

The work in ref. 26 gave a nonlinear optical spectrum filtering effect in double FWM process, where two signal fields could be generated under the conditions of the near two-photon resonance and the single-photon detuning exceeding the band width of absorption of cold atoms. Another point we wanted to point out here was that the off resonance storage in the FWM process was achieved in our system. This result was consistent with work in ref. [27, 28], where Raman quantum memory was performed under low optical depth. In these works, the storage bandwidth was broader than the bandwidth of absorption of cold atoms. Of course, if the detuning of the single-photon was too larger, then the prepared signal could not be stored, agreeing with ref. 29. In ref. 28, single photons with detuning of 200 MHz under the system's maximum limit was stored and retrieved, showing the maximum bandwidth of nonlinear response of media.

In this work, we observed a filtering effect in a FWM process based on an inverted-Y type atomic configuration. We theoretically analyzed and experimentally studied the relation of bandwidth between the input probe and the generated signal, and obtained a conclusion that the high frequency components of edge pulse of generated signal field directly propagated through the media without absorption and the low frequency parts were absorbed. In addition, we observed the storage of the backward side peak by using Raman transition and retrieved later. We obtained the storage process was divided into the region of nonlinear response of atoms. Our results were important in nonlinear optics and optics storage and gave a deep understanding in interaction between light and matters.

**Acknowledgments**

Thanks for discussions with Professor. ShengWang Du. This work was supported by the National Fundamental Research Program of China (Grant No. 2 011CBA00200)**,** the National Natural Science Foundation of China (Grant Nos. 11174271, 61275115，61435011), the Youth Innovation Fund from USTC (Grant No. ZC 9850320804), and the Innovation Fund from CAS.**References**

Fig. 7 The generated signal field vs different period of the half-Gaussian pulse from 1 μs~6 μs. The powers of probe, pump and coupling were 0.45mW, 2.8mW and 0.5mW.

The work in ref. 26 gave a nonlinear optical spectrum filtering effect in double FWM process, where two signal fields could be generated under the conditions of the near two-photon resonance and the single-photon detuning exceeding the band width of absorption of cold atoms. Another point we wanted to point out here was that the off resonance storage in the FWM process was achieved in our system. This result was consistent with work in ref. [27, 28], where Raman quantum memory was performed under low optical depth. In these works, the storage bandwidth was broader than the bandwidth of absorption of cold atoms. Of course, if the detuning of the single-photon was too larger, then the prepared signal could not be stored, agreeing with ref. 29. In ref. 28, single photons with detuning of 200 MHz under the system's maximum limit was stored and retrieved, showing the maximum bandwidth of nonlinear response of media.

In this work, we observed a filtering effect in a FWM process based on an inverted-Y type atomic configuration. We theoretically analyzed and experimentally studied the relation of bandwidth between the input probe and the generated signal, and obtained a conclusion that the high frequency components of edge pulse of generated signal field directly propagated through the media without absorption and the low frequency parts were absorbed. In addition, we observed the storage of the backward side peak by using Raman transition and retrieved later. We obtained the storage process was divided into the region of nonlinear response of atoms. Our results were important in nonlinear optics and optics storage and gave a deep understanding in interaction between light and matters.

**Acknowledgments**

Thanks for discussions with Professor. ShengWang Du. This work was supported by the National Fundamental Research Program of China (Grant No. 2 011CBA00200)**,** the National Natural Science Foundation of China (Grant Nos. 11174271, 61275115，61435011), the Youth Innovation Fund from USTC (Grant No. ZC 9850320804), and the Innovation Fund from CAS.

**References**

1. A. Sommerfeld, Ann. Phys. (Leipzig) **349**, 177 (1914).
2. L. Brillouin, Ann. Phys. (Leipzig) **349**, 203 (1914).
3. H. Jeong, A. M. C. Dawes, and D. J. Gauthier, Phys. Rev. Lett. **96**, 143901 (2006).
4. P. Pleshko and I. Palocz, Phys. Rev. Lett. **22**, 1201 (1969).
5. J. Aavikscoo, J. Lippmaa, and J. Kuhl, J. Opt. Soc. Am. B **5**, 1631 (1988).
6. J. Aavikscoo, J. Kuhl, and K. Ploog, Phys. Rev. A. **44**, R5353 (1991).
7. S.-H. Choi and U. L. Österberg, Phys. Rev. Lett. **92**, 193903 (2004).
8. H. Jeong and U. L. Österberg, Phys. Rev. A. **77**, 021803(R) (2008).
9. Heejeong Jeong, and Shengwang Du, Phys. Rev. A. **79**, 011802(R) 2009
10. William R. LeFew, StephanosVenakides, and Daniel J. Gauthier, Phys. Rev. A. **79**, 063842. 2009
11. Bruno Macke and Bernard Ségard, Phys. Rev. A. **80**, 011803(R) 2009.
12. Dong Wei, J. F. Chen, M. M. T. Loy, G.K. L. Wong, and Shengwang Du, Phys. Rev. Lett. **103**, 093602 (2009)
13. J. F. Chen,HeejeongJeong, L. Feng, M. M. T. Loy, G. K. L. Wong, and Shengwang Du, Phys. Rev. Lett. **104**, 223602 (2010).
14. Shanchao Zhang, J. F. Chen, Chang Liu, M. M. T. Loy, G.K. L. Wong, and Shengwang Du, Phys. Rev. Lett. **106**, 243602 (2011)
15. Shengwang Du, ChinmayBelthangady, PavelKolchin, G. Y. Yin, and S. E. Harris, Optics Letters. **33**. 2149 (2008)
16. Lukin, M.D.; Hemmer, P.R.; Scully, M.O. Adv. At. Mol. Opt. Phys. **42**, 347–386 (2000).
17. Lukin, M. D., Hemmer, P., Loeffler, M., and Scully, M. O., (1998). Phys. Rev. Lett. **81**,2675
18. Hemmer, P. R., et al. Optics Letters. **20**, 982. (1995)
19. Andrew J. Merriam, S. J. Sharpe, H. Xia, D. Manuszak, G. Y. Yin, and S. E. Harris. Optics Letters. **24**, 625 (1999)
20. A. G. Radnaev, Y. O. Dudin, R. Zhao, H. H. Jen, S. D. Jenkins, A. Kuzmich and T. A. B. Kennedy, Nature Physics, **6**, 894-899, (2010).
21. P. S. Hsu, A. K. Patnaik, and G. R. Weich, Optics Letters. **33**, 381 (2008).
22. R. T. Willis, F. E. Becerra, L. A. Orozco, and S. L. Rolston, Phys. Rev. A. **79**, 033814 (2009).
23. Dong-Sheng Ding, Zhi-Yuan Zhou, Bao-Sen Shi, Xu-Bo Zou, and Guang-Can Guo, Phys. Rev. A. **85**, 053815 (2012)
24. A. Gogyan. Phys. Rev. A. **81**, 024304 (2010)
25. Y. Liu, J.-H, Wu., B.-S, Shi., and G.-C, Guo., Chin. Phys. Lett. **29**, 024205 (2012)
26. Yang Liu, Jinghui Wu, Dongsheng Ding, Baosen Shi and GuangcanGuo. New Journal of Physics **14**, 073047 (2012).
27. Dong-Sheng Ding, Wei Zhang, Zhi-Yuan Zhou, Shuai Shi, Guo-Yong Xiang, Xi-Shi Wang, Yun-Kun Jiang, Bao-Sen Shi, Guang-Can Guo. arXiv:1404.0439. (2014)
28. Dong-Sheng Ding, Wei Zhang, Zhi-Yuan Zhou, Shuai Shi, Bao-Sen Shi, Guang-Can Guo. arXiv:1410.7101. (2014)
29. J. H, Wu, D. S, Ding, Y, Liu., Z, Y, Zhou, B. S, Shi, X. B, Zou, and G. C, Guo, Phys. Rev. A. **87**, 013845 (2013)